# Generalizations of a nonlinear fluid model for void formation in dusty plasmas

C S Ng[1], A Bhattacharjee[1], S Hu[2], Z W Ma[3] and K Avinash[4]

[1]Space Science Center, Institute for the Study of Earth, Oceans, and Space, University of New Hampshire, Durham, NH 03824
[2]Department of Physics and Astronomy, University of California at Irvine, Irvine, CA 92697
[3]Institute of Plasma Physics, CAS, Hefei, Anhui 230031, China
[4]Department of Physics and Astrophysics, University of Delhi, Delhi-7, India

Short title: Void formation in dusty plasmas

PACS: 52.27.Lw, 52.30.Ex, 52.35.Mw, 52.20.-j

**Abstract**

New developments in the theory and numerical simulation of a recently proposed one-dimensional nonlinear time-dependent fluid model [K. Avinash, A. Bhattacharjee, and S. Hu, Phys. Rev. Lett. **90**, 075001 (2003)] for void formation in dusty plasmas are presented. The model describes an initial instability caused by the ion drag, rapid nonlinear growth, and a nonlinear saturation mechanism that realizes a quasi-steady state containing a void. The earlier one-dimensional model has been extended to two and three dimensions (the latter, assuming spherical symmetry), using a more complete set of dynamical equations than was used in the earlier one-dimensional formulation. The present set of equations includes an ion continuity equation and a nonlinear ion drag operator. Qualitative features of void formation are shown to be robust with respect to different functional forms of the ion drag operator.

## 1. Introduction

Several dusty (or colloidal) plasma experiments, in laboratory as well as under microgravity conditions, have shown the spontaneous development of voids [1-5]. A void is typically a small and stable centimeter-size region (within the plasma) that is completely free of dust particles and characterized by sharp boundaries. In the laboratory [2], the void is seen to develop from a uniform dust cloud as a consequence of an instability when the dust particle has grown to a sufficient size.

In order to account for their experimental results, Samsonov and Goree [2] suggested that the ion drag force plays a crucial role in causing the initial instability, which can be described as follows. Imagine a local depletion of negatively charged dust particles within a spatially uniform dusty plasma. The depletion will produce a positive space charge with respect to the surrounding plasma, and, hence, an electric field that points outward from the region of reduced dust density. This electric field will cause an inward electrical force, $F_e$, on negatively charged dust particles that tends to restore the dust density to its equilibrium value, and an outward force, $F_d$, due to the ion drag (in the direction of the ion flow) that tends to expel dust particles from the region of depletion. If $F_d > F_e$, which occurs when dust particles have grown to a sufficient size, an instability grows, deepening the initial density depletion.

Stimulated by Ref. [2], there have been essentially two types of theoretical studies: linear stability analyses that include the effect of the ion drag and other effects [6-11], and nonlinear but steady-state analyses that yield void solutions [12-15]. It is of great interest to bridge these two types of theoretical studies by a nonlinear time-dependent model that describes the spontaneous development of the linear instability as well as its subsequent saturation in the nonlinear regime to produce a void.

Recently, we have proposed such a time-dependent, self-consistent nonlinear fluid model for void formation [16-19]. The model is basic and contains three elements: (a) an initial instability caused by the ion drag force $F_d$, (b) a nonlinear saturation mechanism for the instability, and (c) the void as one of the possible nonlinearly saturated states, dynamically accessible from the initially unstable equilibrium. For the initial instability, we choose a simple variant of the zero-frequency mode described by D'Angelo [6]. The saturation mechanism for the instability relies on a crucial nonmonotonic property of the ion drag force that appears to be a robust feature of the force, independent of the regime of collisionality. In the collisional regime, $F_d$ initially increases with the ion velocity $v_i$, attains a maximum for $v_i = v_{thi}$, where $v_{thi}$ is the ion thermal velocity, and decreases for $v_i > v_{thi}$ [20-22], see also discussion in [12, 23]. As the linear instability grows, the ions are initially accelerated in the growing electric field, and $F_d$ initially increases. Eventually, as the ions are accelerated to speeds larger than the ion thermal speed, $F_d$ decreases to balance the electric force $F_e$ and thus saturate the instability.

In Ref. [16], the mechanism for void formation has been studied by analysis and numerical simulation of a simplified set of model equations. The model discussed in Ref. [16] retains only the most essential physical elements needed to illustrate the physics of void formation. Some of the major simplifying assumptions, used in Ref. [16], are: (i) the calculations are carried out in one dimension (1D), (ii) the nonlinear convective term is ignored in the dust fluid momentum equation, (iii) the ion density is assumed to be constant, and (iv) a simple empirical form for the ion drag force is used. The objective of this paper is to present new results, with these simplifications removed.

Before we describe our own results, we mention other studies of void formation. A dust void has been alternately described as a Bernstein-Greene-Kruskal [24] (BGK) mode [15, 25, 26], or a 3D BGK mode in a multi-species plasma [27]. However, such studies assume a completely collisionless plasma, even for dust particles, which is probably not realistic. A minimum-energy model has also been proposed [28]. However, the model depends on a confining potential that is not always present in void experiments. A recent two-dimensional numerical simulation [29] has also been attempted, but it is concluded that the simulation results cannot explain the appearance of the void in the microgravity experiment [3].

Recent experimental results from the PKE-Nefedov facility onboard International Space Station have identified new features of the void formation, such as dust flow vortices around the void [30]. Theoretical and computational research attempting to explain such observations under microgravity has produced interesting results [31-32]. Ref. [32] has also studied the differences in void formation by using different of ion drag models, see also [33].

The following is the outline of this paper. In Section 2, we first show 1D simulations with the convective nonlinearity retained in the dust momentum equation [17]. As in our earlier study, we demonstrate the emergence of a void solution in the saturated state, with a boundary that becomes sharper due to the inclusion of the convective nonlinearity, and present a simple analysis demonstrating how such a sharp solution can be supported by the nonlinear convective term. In Section 3, we present two-dimensional (2D) direct simulations of the model equations, demonstrating that axisymmetric void solutions emerge as relaxed states without any *a priori* assumption of axisymmetry in the dynamics [17,18]. This motivates the extension of our dynamical model to three dimensions (3D) in Section 4, assuming spherical symmetry, which effectively reduces the 3D problem to a 1D radial problem [18]. In Section 5, we present void simulations using a more realistic operator for the ion drag force [21]. We show that the qualitative features of the dynamical relaxation to the void solution remain robust with respect to changes in the functional form of the nonlinear drag operator [18, 19]. We conclude with a summary and a discussion of the implications of our results in Section 6.

## 2. Effect of nonlinear dust convection

We begin with the 1D fluid model equations developed in Ref. [16] in dimensionless form:

$$\frac{\partial v_d}{\partial t} + v_d \frac{\partial v_d}{\partial x} = F_d - E - \alpha_0 v_d - \frac{\tau_d}{n_d} \frac{\partial n_d}{\partial x}, \tag{1}$$

$$F_d = F_{dM} = aE/\left(b + |v_i|^3\right), \tag{2}$$

$$v_i = \mu E, \tag{3}$$

$$\frac{\partial n_d}{\partial t} = -\frac{\partial (n_d v_d)}{\partial x} + D_0 \frac{\partial^2 n_d}{\partial x^2}, \tag{4}$$

$$\frac{\partial n_e}{\partial x} = -\frac{n_e E}{\tau_i}, \tag{5}$$

$$\frac{\partial E}{\partial x} = 1 - n_d - n_e. \tag{6}$$

The notation is identical to that in Ref. [16]. For ease of reference, we summarize here the normalization factors. The spatial coordinate $x$ is in units of $\lambda_{Di} T_e / T_i$ (where $T_{i,e,d}$ is the temperature, $n_{i,e,d}$ is the density, and $m_{i,e,d}$ is the mass of ions, electrons and dust particles respectively, and $\lambda_{Di} = \left[T_i / 4\pi e^2 n_i\right]^{1/2}$ is the ion Debye length), time $t$ is in units of $\omega_{pd}^{-1}$ (where $\omega_{pd} = \left[4\pi n_d Z_d^2 e^2 / m_d\right]^{1/2}$ is the dust plasma frequency with $Z_d$ being the number of electron charge on a dust particle, assumed to be a constant for all particles and independent of time for simplicity), ion fluid velocity $v_i$ is in units of ion thermal speed $v_{thi} = \left[T_i / m_i\right]^{1/2}$, the dust fluid velocity $v_d$ is in units of $\lambda_{Di}T_e\omega_{pd}/T_i$, the electric field $E$ is in units of $T_e/e\lambda_{Di}$, and the ion drag force $F_d$ is in units of $\lambda_{Di}m_dT_e\omega_{pd}^2/T_i$. Note also that $\alpha_0$ is the normalized dust-neutral collision frequency, $\tau_i = T_i/T_e$, and $\tau_d = T_dT_i/T_e^2Z_d$. Electron and ion number densities are measured in units of $n_i$, which is assumed to be constant for simplicity. Note that this assumption is relaxed in our 2D simulations presented later in Section 3 by including an ion equation of continuity in Eq. (11). However, the 2D simulation results show that ion density turns out to be highly uniform anyway. Thus, in the system of equations above, we set $n_i = 1$. The dust number density is in units of $n_i/Z_d$.

In Eq. (1) for the dust fluid velocity $v_d$, the second term in the left-hand-side is the nonlinear convective term, neglected earlier in Ref. [16], for simplicity. The terms in the right-hand-side of Eq. (1) are forces due to the ion drag, the electric field, ion-neutral collisions, and the dust fluid pressure, respectively. Note that the use of a fluid model for dust particles is not valid when the dust particles crystallize and the system develops strong symmetries. However, in a number of void experiments, the dusty plasma appears to be at least partially in a liquid-like colloidal state in the presence of instabilities [1-5]. Indeed, previous theoretical studies [6-12] on the relevant instabilities have relied on fluid models, a practice we continue in the present paper.

In Ref. [16], the ion drag force is modeled by the heuristic form given in Eq. (2), with two positive constants $a$ and $b$. This functional form is motivated by the physical consideration that the ion drag force is proportional to the electric field when the ion $v_i$ fluid velocity is small, but tends to zero as $v_i$ becomes much larger. Equation (3) is the model equation used to relate $v_i$ to the electric field $E$, where $\mu$ is assumed to be a constant mobility coefficient for simplicity. This assumption is removed in Section 5 when we discuss the effect of a more realistic ion drag operator. By comparing with a set of parameters relevant to experiments, we see that the assumption that $v_i$ is linearly proportional to $E$ is acceptable within the range where a void will form, see Fig. 10. Equation (4) is the continuity equation for the dust number density $n_d$, where the last term in the right-hand-side is a diffusion term added primarily for reasons of numerical stability. Equation (5) represents the assumption that the electron fluid is kept in equilibrium by balancing the electric force with the electron fluid pressure. Equation (6) is Gauss' law, enabling us to solve for the electric field $E$ from the charge density.

A simple equilibrium solution for this set of equations corresponds to uniform dust and electron densities (with total charge neutrality) and zero electric field and fluid velocities. A simple analysis outlined in Ref. [16] shows that this equilibrium can be unstable if the condition $a > b$ is satisfied. Following the numerical procedure in Ref. [16], starting from an initial equilibrium with a small dust density modulated by small perturbations, Eqs. (1)-(6) can be integrated forward in time, until the solution saturates into a second steady-state equilibrium with a dust void.

Figure 1 shows a plot of the dust density $n_d$ at position $x = 5$ (solid) and $x = 3$ (dashed) as functions of time, from a simulation for the case with parameters $a = 7.5$, $b = 1.6$, $\alpha_0 = 2$, $\tau_d = 0.001$, $\mu = 1.5$, $\tau_i = 0.125$, and $D_0 = 0.01$. These parameters are chosen to be the same as in simulations presented in Ref. [16]. In Section 5, we will see that the model ion drag based on these parameters has very similar form as that from a more realistic ion drag operator. We see that $n_d(5)$ initially grows nearly exponentially, and then saturates at the level $n_d(5) \sim 1$, while $n_d(3)$ decays slowly in time. In fact when the simulation continues for a much longer time, the dimensionless variable $n_d(3)$ decays to a very small value (approximately $10^{-30}$ or smaller for this set of parameters). Therefore, a final steady state is indeed reached with almost zero dust density within a certain region, referred to earlier as a void solution.

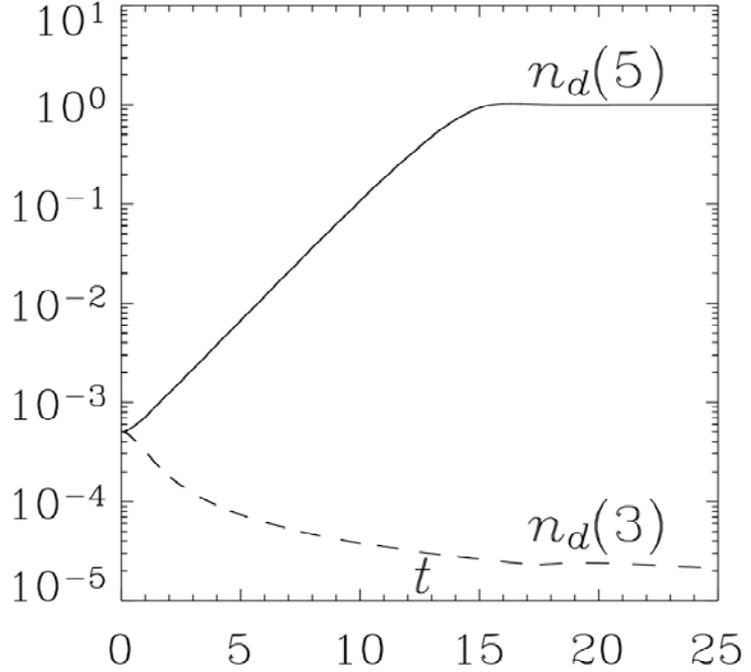

Figure 1. Plot of dust density $n_d$ at $x = 5$ (solid) and $x = 3$ (dashed) as functions of time, from a simulation of Eqs. (1)-(6) with parameters $a = 7.5$, $b = 1.6$, $\alpha_0 = 2$, $\tau_d = 0.001$, $\mu = 1.5$, $\tau_i = 0.125$, and $D_0 = 0.01$.

Figure 2 shows plots of (a) dust fluid velocity $v_d$, (b) dust density $n_d$, (c) electric field $E$, and (d) electron density $n_e$ for such a saturated steady state. The dotted curves are the solutions without the dust convective term [16]. The solid curves are obtained with the same parameters, but with the dust convective term included. We see that the two sets of curves are qualitatively similar, confirming the fact that the nonlinear dust convective term is not required for the realization of a void solution [16]. However, when we examine the plot more closely, we observe that the dust density gradient is sharper (or, the void has a more well-defined boundary) when the dust convective term is included. Note that the dust fluid velocity $v_d$ is nonzero mainly within the void where the dust density $n_d$ is almost zero. In reality, if there is exactly no dust within the void, $v_d$ in that region is actually not well defined, and experimentally non-detectable. Note however that the $v_d$ profile does tend to zero across the void boundary that matches onto a region where the dust density is nonzero.

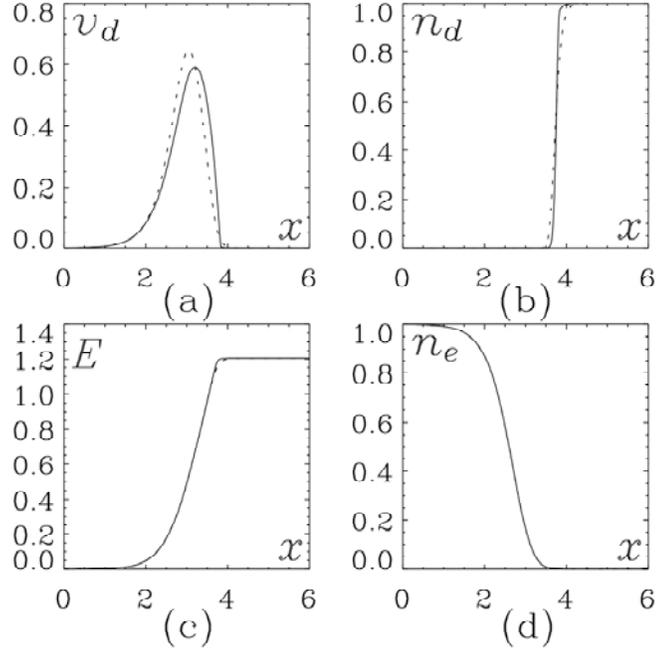

Figure 2. Plots of (a) dust fluid velocity $v_d$, (b) dust density $n_d$, (c) electric field $E$, and (d) electron density $n_e$ for the saturated steady state solution from the simulation with the same parameters as in Fig. 1. Solid curves are for the case with the dust convective term, while the dotted curves are for the case without.

The electron density $n_e$ in this case decays to an unrealistically small value in this 1D calculation, according to Eq. (5), outside the void region. However, in the more realistic solutions presented in Section 3 and 4, $n_e$ does not decay as fast and has a more realistic value outside the void (see, for instance, Fig. 7). A more self-consistent approach would be to allow the dust particles to have variable charge and add a model equation for the charging of the dust particles, as well as more realistic boundary conditions. Since this requires adding substantial complexity to the model, we leave it to a future investigation.

We now show that an infinitely sharp void boundary can only be realized from Eq. (1) if the convective term is included. (By infinitely sharp, we mean that the derivative of the density profile tends to infinity at the void boundary.) Let us consider a steady-state void solution when $D_0 = 0$. By Eq. (4), assuming that there is no dust inside the void, we must have $n_d v_d = 0$. This means that either the dust density $n_d$ or the dust fluid velocity $v_d$ has to be zero at all spatial locations. An infinitely sharp void boundary at $x = x_v$ will then have $n_d^- = n_d(x_v - 0^+) = 0$, and $n_d^+ = n_d(x_v + 0^+) \neq 0$, or formally $n_d = n(x)[1 - \Theta(x + x_v) + \Theta(x - x_v)]$, where $\Theta(x)$ is the Heaviside step function such that $d\Theta(x)/dx = \delta(x)$. This formal solution describes a dust void with a sharp boundary at $x = \pm x_v$. From the steady-state form of Eq. (1), this sharp-boundary solution will produce an infinitely large force (in the form of a delta-function) at $x = \pm x_v$ due to the presence of the $\partial n_d / \partial x$ term in the dust pressure gradient. We see that if the dust convective term is ignored, there is no other term that can balance this discontinuity in the dust pressure term. However, if we include the convective term, with the dust fluid velocity of the form $v_d = u(x)[\Theta(x + x_v) - \Theta(x - x_v)]$, the delta-function terms can be balanced if $u(\pm x_v)^2 = 4\tau_d$. This demonstrates that the convective term is indeed essential in obtaining an infinitely sharp boundary in the final void solution. In an actual physical experiment, such a sharp boundary will be smoothened by diffusion and dissipation.

Before we conclude this Section, we remark on the necessity of including a numerical diffusion coefficient $D_0$ in the ion continuity equation (4). It is evident from the discussion above that if we set $D_0 = 0$, the relaxed state tends to a solution with an infinitely sharp boundary. Numerical instability will appear due to lack of resolution as the system approaches a steady state if $D_0 = 0$, no matter what the spatial grid size may be. At the end of Section 4, we will present numerical results showing that qualitatively similar void solutions form at even lower values of $D_0$, and that the boundaries of these solutions are sharper.

## 3. Void solutions in two dimensions

In Section 2 and Ref. [16], we have shown that dust void solutions can indeed be produced from an initially unstable equilibrium in a 1D simulation, both with and without the convective nonlinearity. We now extend further the scope of our investigation, and present results from 2D simulations. We first write down the 2D fluid equations, extending the 1D equations (1)-(6):

$$\frac{\partial \mathbf{v}_d}{\partial t} + \mathbf{v}_d \cdot \nabla \mathbf{v}_d = \mathbf{F}_d - \mathbf{E} - \alpha_0 \mathbf{v}_d - \frac{\tau_d}{n_d} \nabla n_d, \tag{7}$$

$$\mathbf{F}_d = \mathbf{F}_{dM} = a\mathbf{E} \Big/ \left(b + |\mathbf{v}_i|^3\right), \tag{8}$$

$$\mathbf{v}_i = \mu \mathbf{E}, \tag{9}$$

$$\frac{\partial n_d}{\partial t} = -\nabla \cdot (n_d \mathbf{v}_d) + D_0 \nabla^2 n_d, \tag{10}$$

$$\frac{\partial n_i}{\partial t} = -\nabla \cdot (n_i \mathbf{v}_i) + A n_i - C n_e, \tag{11}$$

$$\nabla n_e = -\frac{n_e \mathbf{E}}{\tau_i}, \tag{12}$$

$$\nabla \cdot \mathbf{E} = n_i - n_d - n_e. \tag{13}$$

These equations are mostly Eqs. (1)-(6), written in vector form, except that we now include Eq. (11), a continuity equation for the ion density $n_i$, instead of assuming a uniform background.

The two constant coefficients $A$ and $C$ model the processes of ionization and recombination with electrons.

A 2D code has been developed to simulate these equations in cylindrical geometry. The numerical scheme uses a fourth-order Runge-Kutta method in time, with fourth order central finite differencing in the radial direction. We implement a pseudo-spectral method in the azimuthal direction. Open boundary conditions are used in to outer radial positions. As in the 1D simulations, we begin with an equilibrium state in which all densities are uniform, and with small perturbations superposed. The system evolves through a linear exponential phase and a nonlinear near-explosive phase before relaxing to a stable void solution.

Figure 3 shows a contour plot of the dust density $n_d$ in the saturated state exhibiting a void with a sharp boundary, using experimentally representative parameters $a = 100$, $b = 1.6$, $\alpha_0 = 2$, $\tau_d = 0.01$, $\mu = 10$, $\tau_i = 10$, $D_0 = 0.01$, $A = 0.8$, $C = 1$. Despite the fact that the numerical simulation enables radial as well as angular dependencies in the evolution of the dynamics, the final relaxed state turns out to be nearly axisymmetric. This can also be seen from the contour plots of the dust fluid velocity $\mathbf{v}_d$. We see that the radial component $v_{dr}$ (Fig. 4), on the average, has a much larger magnitude than the azimuthal component $v_{d\phi}$ (Fig. 5). Unlike in our previous 1D simulations, the ion density $n_i$ is now allowed to evolve in time through the continuity equation (11). However, it turns out that the ion density is quite uniform in the final state, as shown in Fig. 6, which is also approximately axisymmetric. Since the void solution is nearly axisymmetric, we can see the solution more quantitatively through 1D radial profiles, as shown in Fig. 7 for (a) radial component of the dust fluid velocity $v_{dr}$, (b) dust density $n_d$, (c) radial electric field $E_r$, and (d) electron density $n_e$, as functions of radial position $r$ in a cut through the center (solid traces). From these results, we deduce that the void formation mechanism found in our previous 1D model, also works in 2D.

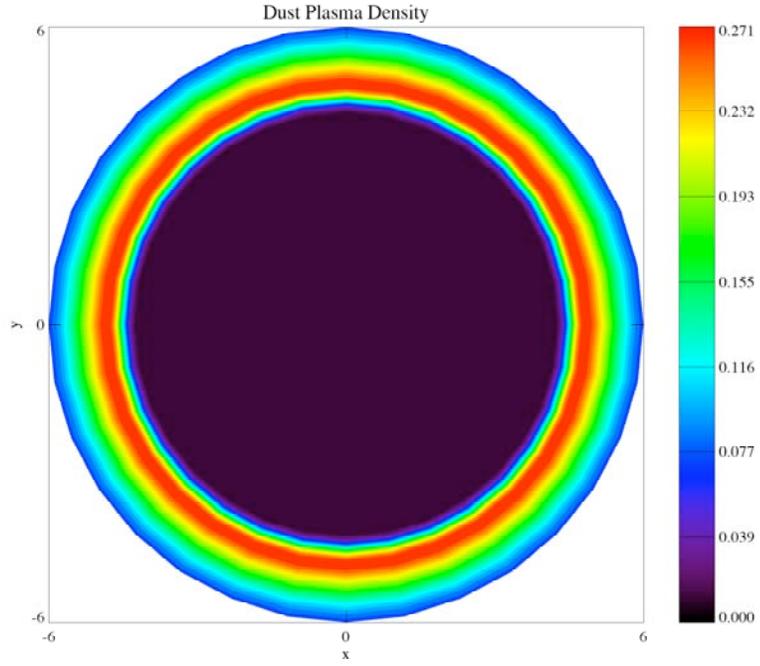

Figure 3. Contour plot of the dust plasma density $n_d$ in the saturated state of a 2D simulation of Eqs. (7)-(13) for parameters $a = 100, b = 1.6, \alpha_0 = 2, \tau_d = 0.01, \mu = 10, \tau_i = 10, D_0 = 0.01, A = 0.8, C = 1$.

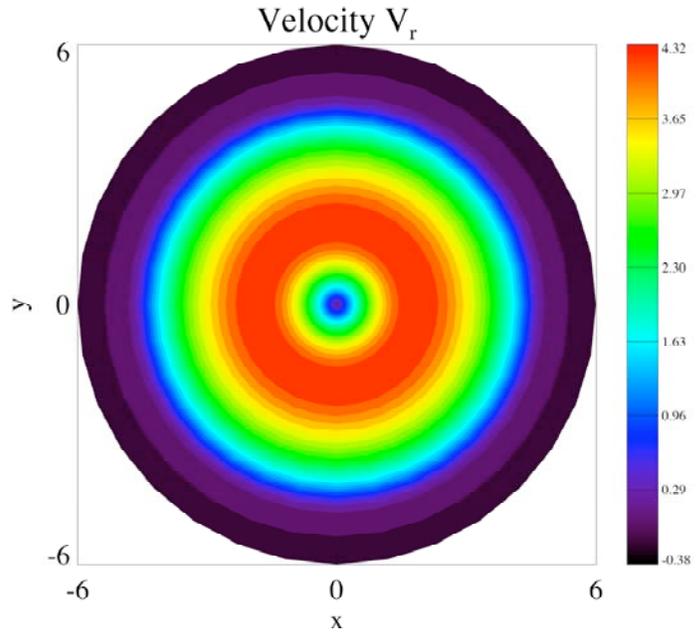

Figure 4. Contour plot of the radial component of the dust fluid velocity $v_{dr}$ for the same saturated solution as in Fig. 3.

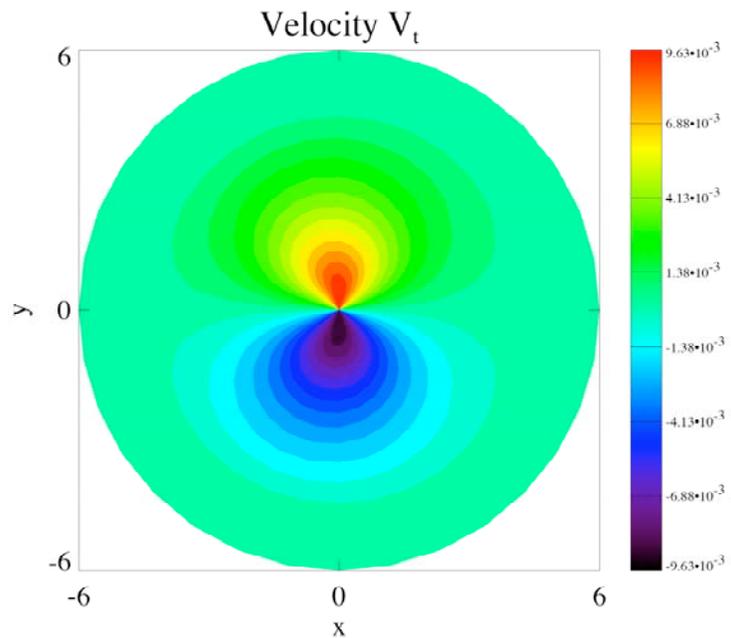

Figure 5. Contour plot of the azimuthal component of the dust fluid velocity $v_{d\phi}$ for the same saturated solution as in Fig. 3

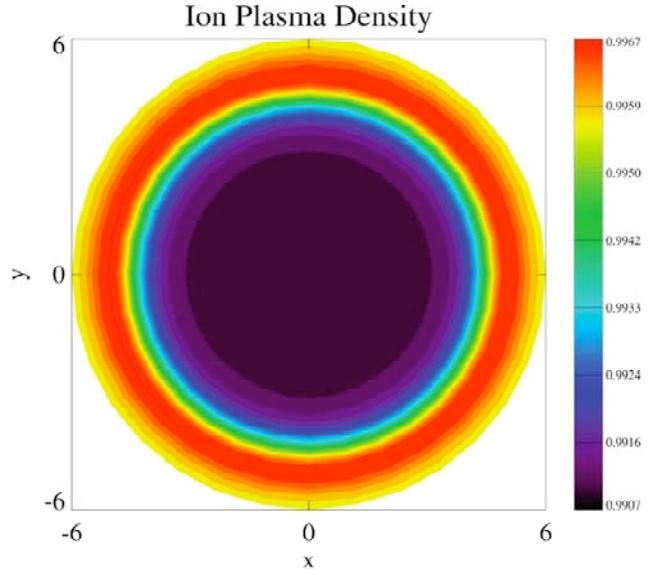

Figure 6.  Contour plot of the ion density $n_i$ for the same saturated solution as in Fig. 3.

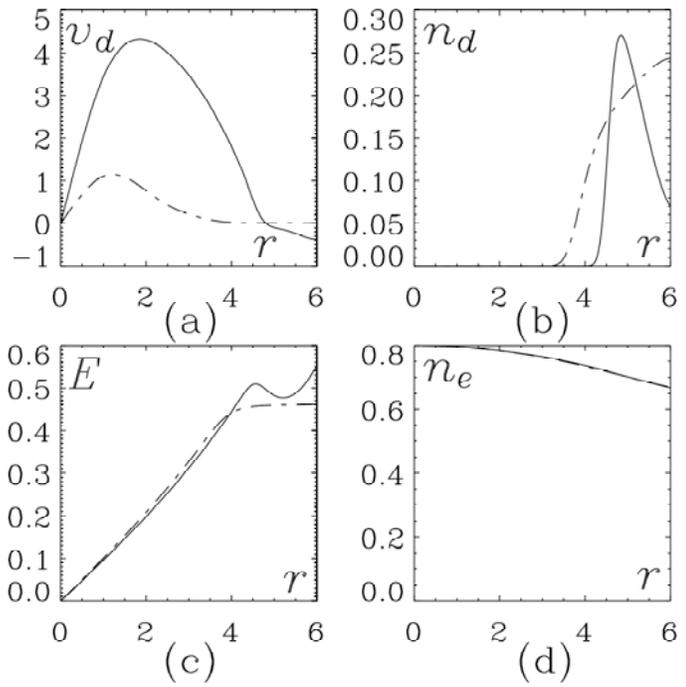

Figure 7.  Solid traces are profiles of (a) radial component of the dust fluid velocity $v_{dr}$, (b) dust density $n_d$, (c) radial electric field $E_r$, and (d) electron density $n_e$, as functions of radial position $r$ in a cross-section of the saturated

solution in Fig. 3 to 6. Dash-dot traces are the corresponding plots from a simulation of Eqs. (14)-(19) for the 2D cylindrically symmetric case ($j = 1$) using the same parameters.

## 4. Three-dimensional symmetric solutions

Now that we have confirmed that our dust void model works in 2D as well as 1D, it is of great interest to extend the solutions to 3D, since voids observed in experiments are actually 3D structures. However, a fully 3D simulation is difficult. Fortunately, we can take a cue from our fully 2D studies, which suggest that the system tends to symmetric solutions even in the presence of symmetry-breaking perturbations. As a first step, we are thus motivated to investigate void formation in 3D, assuming spherical symmetry. It is also instructive to compare the 3D spherically symmetric solutions with the 2D cylindrically symmetric solutions. The radial equations for symmetric solutions in both 2D and 3D are quite similar:

$$\frac{\partial v_{dr}}{\partial t} + v_{dr}\frac{\partial v_{dr}}{\partial r} = F_{dr} - E_r - \alpha_0 v_{dr} - \frac{\tau_d}{n_d}\frac{\partial n_d}{\partial r} , \tag{14}$$

$$F_{dr} = F_{drM} = aE_r \Big/ \left(b + |v_{ir}|^3\right) , \tag{15}$$

$$v_{ir} = \mu E_r \tag{16}$$

$$\frac{\partial n_d}{\partial t} = -\frac{1}{r^j}\frac{\partial(r^j n_d v_{dr})}{\partial r} + D_0 \frac{1}{r^j}\frac{\partial}{\partial r}\left(r^j \frac{\partial n_d}{\partial r}\right) , \tag{17}$$

$$\frac{\partial n_e}{\partial r} = -\frac{n_e E_r}{\tau_i} , \tag{18}$$

$$\frac{\partial E_r}{\partial r} + \frac{jE_r}{r} = 1 - n_d - n_e , \tag{19}$$

with all non-radial components of vectors set to zero. The integer $j$ in (17) and (19) is 1 for the 2D cylindrically symmetric case, and 2 for the spherically symmetric case. Note also that the 1D equations (1) – (6) are recovered by setting $j = 0$. Eqs. (14)-(19) can then be simulated in very much the same way as in the 1D case.

Figure 7 compares the output from a fully 2D simulation (solid) to those from a simulation of the radial equation for the cylindrically symmetric case (dash-dotted), using the same parameters. We see that the two sets of results agree quite well qualitatively, and even quantitatively. The main quantitative different is in the dust fluid velocity $v_{dr}$, and that too mostly in the dust-free region. This is understandable, since when the dust density $n_d$ is close to zero, the value of the dust fluid velocity is not well constrained. Other differences can occur due to small asymmetric effects, as well as the non-constant ion density in the 2D runs.

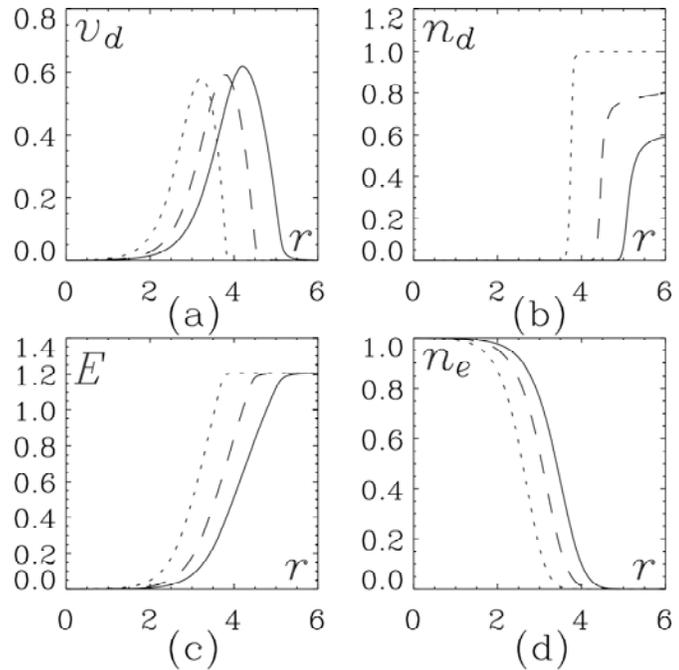

Figure 8.   Plots of (a) dust fluid velocity $v_{dr}$, (b) dust density $n_d$, (c) electric field $E_r$, and (d) electron density $n_e$ for the saturated steady state solution, as functions of radial position $r$, from the simulation of the radial equations for the 3D spherically symmetric case (solid curves), the 2D cylindrically symmetric case (dashed curves), and the 1D case (dotted curves), using the same parameters as in Fig. 1.

Figure 8 shows plots of (a) dust fluid velocity $v_{dr}$, (b) dust density $n_d$, (c) electric field $E_r$, and (d) electron density $n_e$ for the saturated steady state solution, as functions of radial position $r$, from the simulation of the radial equations for the 3D spherically symmetric case (solid curves). For comparison, corresponding plots are also shown for the 2D cylindrically symmetric case (dashed curves), and for 1D case (dotted curves), using the same parameters as in Fig. 1. We see that all three cases give qualitatively similar void solutions, confirming that the void formation model does extend to 3D, assuming spherical symmetry. The main quantitative differences from 1D to the 3D radial case are the increase in the size of the void size and that the dust density attains its asymptotic value over a larger radial distance.

We now return to the issue of the numerical diffusion term in Eqs. (4), (10) and (17), added to avoid numerical instability due to a sharp boundary in the dust density. In Fig. 9, we compare the dust density $n_d$ of the 2D cylindrically symmetric solution (dashed trace, presented earlier as Fig. 7(b)), for which $D_0 = 0.01$. The solid trace represents $n_d$ from a run using identical parameters except with a smaller numerical diffusion coefficient ($D_0 = 0.001$). We see that the two runs have almost identical final saturation states. Upon closer examination, we note that the dust density has a larger gradient (sharper dust boundary) for $D_0 = 0.001$ than for $D_0 = 0.01$. This is consistent with the analytical demonstration that an infinitely sharp dust void should form if $D_0 = 0$.

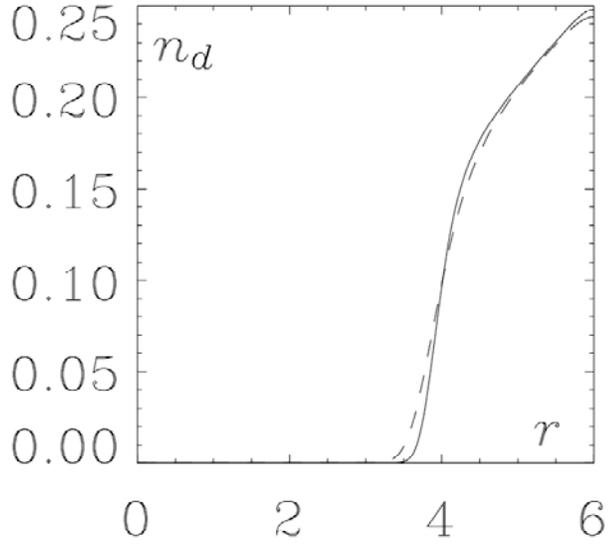

Figure 9.  Dust density $n_d$ of the 2D cylindrically symmetric solution from Fig. 7(b) (dashed trace), with $D_0 = 0.01$. The solid trace is $n_d$ from a run using identical parameters except with a smaller dust diffusion coefficient ($D_0 = 0.001$).

## 5. A more realistic ion drag operator

So far our simulations are based on a model form of ion drag operator, as in (2) or (8). Since the ion drag force plays an important physical role in the process of void formation, it is of great interest to see if voids can still form in our model with a more realistic form of the ion drag operator. To do so, we will use the form recently derived by Khrapak *et al* [21]. After normalizing in the same way as in Section 2, the Khrapak ion drag force operator can be written as,

$$F_d = F_{dK} = \frac{AE}{(1+c_1 E)^{1/2} u} \left\{ \left[ \frac{1}{u} e^{-u^2/2} - \frac{1}{u^2} \sqrt{\frac{\pi}{2}} \text{erf}\left(\frac{u}{\sqrt{2}}\right) \right] \times \right.$$
$$\left. \left[1 + 2z\tau - 4z^2\tau^2 \ln \Lambda \right] + \sqrt{\frac{\pi}{2}} \text{erf}\left(\frac{u}{\sqrt{2}}\right)\left[2(1+z\tau) + u^2\right] + u e^{-u^2/2} \right\}, \quad (20)$$

with

$$u = c_0 E / (1 + c_1 E)^{1/2},$$

$$c_0 = \mu_0 T_e / p v_{thi} e \lambda_{Di}, \quad c_1 = \alpha T_e / p e \lambda_{Di},$$

$$A = \mu_0 T_i a_d^2 n_i \sqrt{2\pi} / p v_{thi} Z_d e,$$

$$\Lambda = (R + \lambda)/(R + a_d), \quad \lambda_{De0}^2 = T_e / 4\pi e^2 n_i,$$

$$\lambda = \left[ \frac{n_e}{\lambda_{De0}^2} + \frac{1}{\lambda_{Di}^2 (1 + u^2)} \right]^{-1/2},$$

$$R = R_0 / (1 + u^2) = Z_d e^2 / T_i (1 + u^2),$$

$$z = Z_d e^2 / a_d T_e, \quad \tau = T_e / T_i,$$

where $\mu_0 = 1460 \text{ cm}^2/\text{Vs}$, $p$ is the pressure in Torr, $\alpha = 0.0264 \text{ cm/V}$, $a_d$ is the radius of the dust particles, and erf denotes the error function.

It is possible to choose these plasma parameters such that the ion drag force $F_{dK}$ given by Eq. (20) is similar qualitatively to the model operator $F_{dM}$, as in Eq. (2). Figure 10 shows plots of the ratio of $F_{dK}/E$ (solid) and ion velocity $u_K$ (dashed) using the Khrapak operator with $T_e = 1 \text{ eV}$, $T_i = 0.125 \text{ eV}$, $Z_d = 8682$, $n_i = 3 \times 10^9 \text{ cm}^{-3}$, $m_i = 6.6 \times 10^{-26} \text{ kg}$, $p = 100$ Pa, $a_d = 3.16 \times 10^{-6}$ m, as functions of $E$. On the same figure, we also over-plot $F_{dM}/E$ (dot) and ion velocity $u_M$ (dash-dotted) using the model operator with $a = 7.5, b = 1.6, \mu = 1.5$. The two arrows indicate values of saturated electric field $E_{sK}$ and $E_{sM}$ when $F_d = E$. We see that for the parameter range considered in this paper, the profile of a more realistic ion drag operator can indeed be approximated reasonably well by the model operator that we have been using so far.

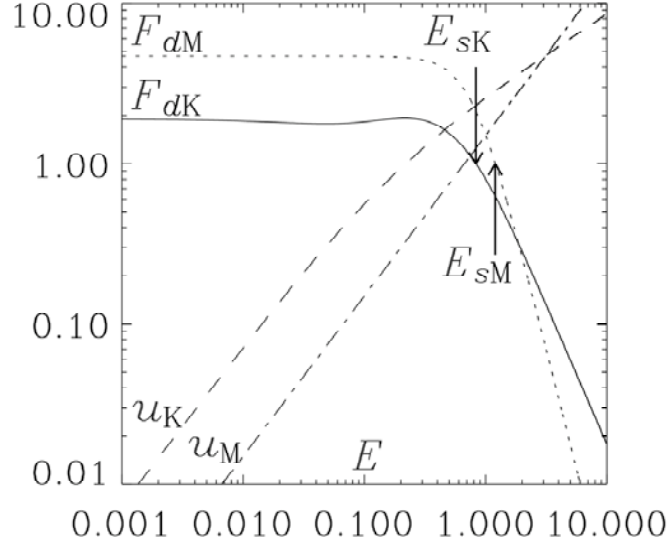

Figure 10. Ratio of $F_{dK}/E$ (solid) and ion velocity $u_K$ (dashed) using the Khrapak operator with $T_e = 1\,\text{eV}$, $T_i = 0.125\,\text{eV}$, $Z_d = 8682$, $n_i = 3 \times 10^9\,\text{cm}^{-3}$, $m_i = 6.6 \times 10^{-26}\,\text{kg}$, $p = 100\,\text{Pa}$, $a_d = 3.16 \times 10^{-6}\,\text{m}$, as functions of $E$, over-plotted with $F_{dM}/E$ (dotted) and ion velocity $u_M$ (dash-dotted) using the model operator with $a = 7.5$, $b = 1.6$, $\mu = 1.5$. The two arrows indicate values of saturated electric field $E_{sK}$ and $E_{sM}$ when $F_d = E$ for the two cases respectively.

We can substitute the Khrapak operator, in the form given by Eq. (20), directly into Eqs. (1), (7), or (14), instead of the model operator specified by Eqs. (2), (8) or (15), and repeat our simulations to test if our previous conclusions hold. Figure 11 shows plots of dust velocity ($v_d$), dust density ($n_d$), electric field ($E$) and electron density ($n_e$) as functions of $r$ (or $x$ for the 1D run), for a saturated void solution using our 1D (solid) and 3D spherically symmetric (dash-dotted) simulations, with $\alpha_0 = 2$, $\tau_d = 0.001$, $\tau_i = 0.125$, and $D_0 = 0.01$, using the Khrapak ion drag operator, with the same parameters as in Fig. 10.

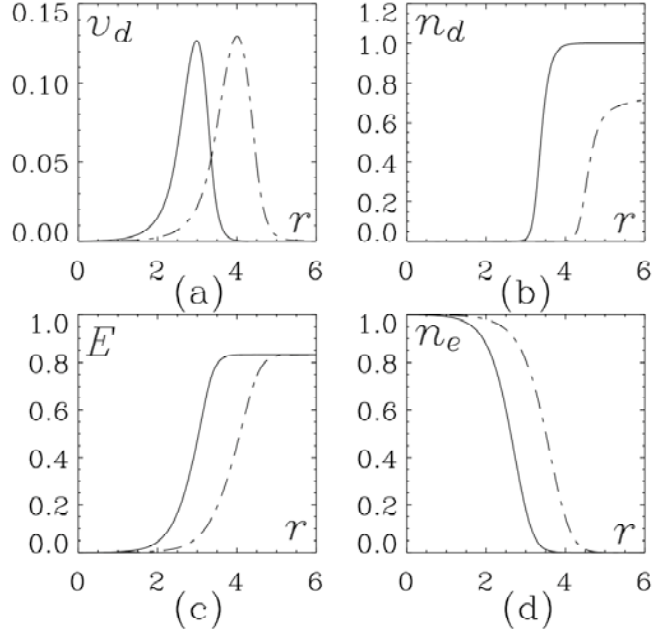

Figure 11. Dust velocity ($v_d$), dust density ($n_d$), electric field ($E$) and electron density ($n_e$) as functions of $r$, for a saturated void solution using our 1D (solid) and 3D spherically symmetric (dash-dotted) simulations, with $\alpha_0 = 2$, $\tau_d = 0.001$, $\tau_i = 0.125$, and $D_0 = 0.01$, using the Khrapak ion drag operator, with the same parameters as in Fig. 10.

We see that a final void state does indeed form, with profiles that are qualitatively similar to those obtained using the model ion drag operator. While this is reassuring, it should be mentioned that there are quantitative differences between the predictions of various operators. These discrepancies do matter in assessing the accuracy of various operators for a specific experiment, and also in determining whether the ion drag force is the dominant mechanism controlling void formation in the experiment. We should also remark that voids do not form in all plasma parameter regimes. For example, in some parameter regimes, the electric field saturates at a sufficiently low magnitude that is not strong enough to sustain a void. We note that a very recent paper has reported experimental results under microgravity conditions that cause the void to close in some parameter regimes [34]. Comparison of our theory with these experimental results is left for a future investigation.

## 6. Conclusion

In this paper, we have shown that the void formation mechanism based on a fluid model proposed in Ref. [16] is robust with respect to generalizations in many aspects. One-dimensional simulations as well as analytical solutions with the dust convective term included have shown that void solutions form with even sharper boundaries. This can be understood by a simple force balance analysis of the dust fluid equation of motion. The basic mechanism for void formation produced by an instability caused by the ion drag force has been shown to work in higher dimensions, first by means of a fully 2D simulation, which shows the tendency to relax to cylindrically symmetric solutions. This, in turn, motivated the extension to 3D spherically symmetric solutions. We have shown that for symmetric cases, dimensionality does not affect the tendency for void formation. There can be, of course, quantitative differences depending on dimensionality, with void structures that are slightly larger in size in the 3D radial case compared with those in 1D. We have also shown that the ion drag force model used so far can give a reasonably close approximation to a more realistic ion drag operator [21], and that voids can still form for experimentally relevant plasma parameter regimes, when such an operator is used in the simulations.

Since this model is still based on many simplifying assumptions, we do not claim that it is complete or that it can explain all the experimental details of void formation. There are several possible ways in which the basic model can be improved, such as the inclusion of full equations of motion for ions as well as electrons, the effect of a neutral fluid, the ionization and recombination processes, the charging of dust particles with variable charge and radius, strong coupling effects between dust particles, kinetic effects, and experimentally realistic geometry and boundary conditions. All these improvements will require much more effort, and is best done in a step-by-step manner that will enable us to understand their physical consequences for our model. The results presented in this paper represent some major improvements on the basic time-dependent nonlinear model presented in Ref. [16]. With respect to these improvements, we have shown that the fluid model is quite robust in demonstrating the formation of dust voids. Further extensions, along the lines suggested above, are left to future work.

**Acknowledgments**

This research is supported by the Department of Energy Grant No. DE-FG02-04ER54803.